\documentclass[12pt]{article}
\usepackage{amsfonts,amsmath}
\usepackage{amssymb}

 \makeatletter \@addtoreset{equation}{section} \makeatother

\newcommand{\etc}{{\it etc}}

\newcommand{\PH}{{  {PH} }}
\newcommand{\HG}{H_{M} }
\newcommand{\HGC}{H_{M}( {\mathbb{C}}) }

\newcommand{\FFf}{\mathcal{F}}
\newcommand{\ff}{\large{\texttt{f}}}

\renewcommand{\Re}{\mathop{\rm Re}}
\renewcommand{\Im}{\mathop{\rm Im}}

\newcommand{\cY}{{\overline{\gY\rule{0pt}{10pt}}}}
\newcommand{\gY}{{\mathcal{Y}}}

\newcommand{\W}{\mathcal{W}}

\newcommand{\za}{{{A}}}
\newcommand{\un}{{{n}}}
  \newcommand{\zb}{{{B}}}

\newcommand{\QQ}{\mathbf{Q}}

 \newcommand{\hhh}{\nu^{-1}}%
 \newcommand{\hhhh}{ \nu}%

 \newcommand{\be}{\begin{equation}}
\newcommand{\ee}{\end{equation}}
\newcommand{\bee}{\begin{eqnarray}}
\newcommand{\beee}{\begin{array}}
\newcommand{\eee}{\end{eqnarray}}
\newcommand{\eeee}{\end{array}}

\newcommand{\gn}{\nu}
\newcommand{\gm}{\mu}
\newcommand{\gc}{\chi}
\newcommand{\gx}{\xi}
\newcommand{\gf}{\zeta}
\newcommand{\gr}{\rho}
\newcommand{\ga}{\alpha}
\newcommand{\gb}{\beta}
\newcommand{\gga}{\gamma}

\newcommand{\M}{{\cal M}}

\newcommand{\A}{{\cal A}}

\newcommand{\B}{G}
\newcommand{\Bb}{{\cal B}}
\newcommand{\F}{{\cal F}}
\newcommand{\E}{{\cal E}}
\newcommand{\ls}{\!\!\!\!\!\!}
\newcommand{\htimes}{\subset{\ls \times}}
\newcommand{\hltimes}{\subset{\!\!\!\!  \times}}

\newcommand{\Z}{{{\cal Z}}}
\newcommand{\cZ}{{\overline{{\cal Z}}}}

\newcommand{\dis}{\displaystyle}

\newcommand{\gd}{\delta}

\newcommand{\go}{\omega}

\newcommand{\by}{{\bar{y}}}

\newcommand{\q}{\,,\qquad}
\newcommand{\ie}{{\it i.e.,} }

\newcommand{\da}{{\alpha^\prime}}
\newcommand{\db}{{\beta^\prime}}

\newcommand{\ZIGM}{\mathfrak{H}_M}
\newcommand{\GZM}{\mathbb{C}^M\times\mathfrak{H}_M}

\newcommand{\X}{{ \mathbf{ X}}}
\newcommand{\Y}{{\mathbf{Y}}}

\newcommand{\bZ}{{\overline{Z}}}
\newcommand{\nn}{\nonumber}
\newcommand{\half}{\frac{1}{2}}

\newcommand{\p}{\partial}

\renewcommand{\P}{{\cal P}}
\newcommand{\Hh}{{ \cal H}}

\newcommand{\D}{{\cal D}}

\newcommand{\C}{{\cal C}}

\newcommand{\f}{\frac}
\newcommand{\TT}{   \mathrm{T}}
\newcommand{\SSS}{   \mathrm{S}}
\newcommand{\G}{{\tilde \xi}}

\newcommand{\rank}{\mathop{\it rank}\nolimits}

\renewcommand{\G}{\mathcal{G}}

\renewcommand{\gn}{\mathbf{E}}
\renewcommand{\gm}{\mathbf{D}}
\renewcommand{\gc}{{H}}
\renewcommand{\gx}{{Q}}
\renewcommand{\gf}{{F}}
\renewcommand{\gr}{{R}}
\renewcommand{\ga}{{A}}
\renewcommand{\gb}{{B}}

\usepackage{hyperref}

\begin{document}

\begin{flushright}
\vspace{1mm}
FIAN/TD/02-09\\ January {2009}\\
\end{flushright}

 \begin{center}
 {{\large\bf $Sp(8)$ invariant higher spin theory, twistors
 and   geometric\\ BRST formulation of unfolded field equations}}
 \vglue 0.6  true cm

 O.A. Gelfond$^1$ and M.A.~Vasiliev$^2$
 \vglue 0.3  true cm

 ${}^1$Institute of System Research of Russian Academy of Sciences,\\
 Nakhimovsky prospect 36-1,
 117218,
 Moscow, Russia
 \vglue 0.3  true cm

 ${}^2$I.E.Tamm Department of Theoretical Physics, Lebedev Physical
 Institute,\\
 Leninsky prospect 53, 119991, Moscow, Russia
 \end{center}

 \begin{abstract}
We discuss twistor-like interpretation of the $Sp(8)$ invariant
formulation of $4d$ massless fields in ten dimensional Lagrangian
Grassmannian   $Sp(8)/P$ which is the generalized space-time in
this framework. The  correspondence space $\mathbf{C}$ is
 $SpH(8)/PH$ where $SpH(8)$ is the semidirect product of $Sp(8)$
 with Heisenberg group $\HG$ and $PH$ is some
quasiparabolic subgroup of $SpH(8)$. Spaces of
functions on  $Sp(8)/P$ and $SpH(8)/PH$ consist of
$Q_P $ closed functions on $Sp(8)$ and $Q_{PH} $ closed functions on
$SpH(8)$, where $Q_P $ and $Q_{PH}$ are canonical BRST operators
of $P$ and $PH$. The space of functions on the generalized twistor
space $\mathbf{T}$ identifies with the $SpH(8)$ Fock module. Although $\mathbf{T}$
cannot be realized as a homogeneous space, we find a
nonstandard $SpH(8)$ invariant BRST operator $\QQ$ $(\QQ^2 =0)$ that
gives rise to an appropriate class of functions  via the
condition $\QQ f=0$   equivalent to the unfolded  higher--spin
equations. The proposed construction is manifestly $Sp(8)$ invariant,
globally defined and coordinate independent. Its Minkowski analogue
 gives  a version of twistor theory with both types of chiral spinors
 treated on equal footing.
The extensions to the higher rank case with several  Heisenberg
groups and to the complex case are considered.
A relation with  Riemann theta
functions, that are $\QQ$-closed, is  discussed.

 \end{abstract}

\newpage
\tableofcontents
\newpage

 \section{Introduction}\label{Generalities}

In \cite{F} Fronsdal suggested that a set of massless fields of all
spins in the four dimensional space should possess a natural
description in the $Sp(8)$ invariant ten dimensional Lagrangian
Grassmannian \, $ Sp(8|\mathbb{R})/P$ described by the  diagram $\,\,\,\,
\begin{picture}(90,4)(0,6)
{\linethickness{.25mm}
\put(0,9){\line(1,0){60}}%
\put(60,8){\line(1,0){30}}%
\put(60,10){\line(1,0){30}}%
\put(81.5,5.5){{ \large $\mathbf \times$}} \put(70,5.5){{ \large
$\mathbf <$}}
 \put(-5,6.3)  {{  \bf $\bullet$}}
\put(25,6.3) {{ \bf $\bullet$}}
\put(55,6.3) {{ \bf $\bullet$}} }
\end{picture}\,
\,\,,$ that results from crossing out the right node of the Dynkin
diagram  of $Sp(8 )$ thus indicating which parabolic subgroup of
$Sp(8)$  should be chosen in the quotient space construction
\cite{Bast_East}. The big cell of $\C \M_M= Sp(2M|\mathbb{R})/P$ we denote
$\M_M$ for any $M$. Local coordinates of
$\M_M$ are real symmetric matrices $X^{AB}=X^{BA}$ ($A,B, \ldots =
1,\ldots M$). The conclusion that $4d$ massless fields can be
described in $\M_4$ was also reached in \cite{BLS1}.

The dynamical equations in $\M_M$, that for $M=4$ are equivalent to
the field equations for massless fields of all spins in the
four dimensional Minkowski space
, were obtained in \cite{BHS}
by using so-called unfolded formulation of the massless field
equations \cite{4dun, Ann, un, act}
\be \label{dydy} \left ( \f{\p}{\p X^{ A B}} + \mu 
\f{\p^2}{\p Y^ A \p Y^ B}\right ) C(Y|X) =0\,, \ee where $Y^ A$
were treated as auxiliary commuting variables and $\mu$ is an
arbitrary non-zero parameter introduced for future convenience. It
should be stressed that the infinite systems of $4d$ massless
fields of all spins $s=0,1,2,\ldots$ and $s=1/2,3/2,\ldots $
described by (\ref{dydy}) are particularly interesting as
constituting fundamental multiplets of fields that appear in the
$4d$ nonlinear theories of higher-spin  massless fields (see
\cite{Gol} and references therein).

The equations (\ref{dydy}) express the first derivatives with
respect to space-time coordinates $X^{AB}$ in terms of the fields
themselves. As such, they belong to the class of  {\it unfolded}
partial differential equations  that, more generally, express the
exterior differential of a set of differential forms in terms of
exterior products of the differential forms themselves (for more detail
see Section \ref{UnfoTwist}). Such a
first-order form of dynamical field equations can always be achieved
by introducing a (may be infinite) set of auxiliary fields which
parametrize all combinations of derivatives of the dynamical fields
that remain unrestricted by the field equations.

For example, in the
system (\ref{dydy}), the {\it dynamical fields} are \be\label{b0}
b(X)=C(0|X) \ee and \be \label{f0}
 f_A(X)= \f{\p}{\p Y^A} C(Y|X)\Big |_{Y=0}\,.
\ee

As a consequence of (\ref{dydy}), they satisfy, respectively, the
equations \be \label{oscal} \Big ( \f{\p^2}{\p X^{ A B} \p X^{ C D}}
- \f{\p^2}{\p X^{ A C} \p X^{ B D}}\Big ) b(X)=0\, \ee
and
\be \label{ofer}
\f{\p}{\p X^{ A B}} f_ C(X) - \f{\p}{\p X^{ A C}} f_ B(X)      =0\,.
\ee
 All the fields
 \be
 \label{gener}
C_{A_1\ldots A_n} (X) = \f{\p^n}{\p Y^{A_1}\ldots \p Y^{A_n}}
C(Y|X)\Big |_{Y=0} \q n>1 \ee are auxiliary, being expressed via
higher $X$--derivatives of the dynamical fields by virtue of the
equations (\ref{dydy}). In  \cite{BHS} it was shown that the
equations (\ref{oscal}) and (\ref{ofer}) along with constraints that
express the auxiliary fields via $X$--derivatives of the dynamical
fields exhaust the content of the unfolded system (\ref{dydy}). That
the equations (\ref{dydy})  formulated in the ten dimensional
space-time still describe massless fields in four dimensions was
also shown in \cite{BHS} using the unfolded dynamics approach.

The description of infinite towers of $4d$ massless fields in $\M_M$
is compact, efficient and have a number of remarkable properties.
The related issues have been studied in a number of papers
\cite{Mar,cur,IB,DV,PST,tens2,BPST,BBAST,EL,west,EI,33,gelcur}.
In our recent paper \cite{gelcur} the form of the unfolded field
equations for conserved higher-spin currents was  worked out. In particular,
it was shown that the consistent formulation of currents requires
extension of the equations (\ref{dydy}) to complex coordinates
\be\label{Sieg}
 \Z^{AB} =X^{AB}+i\,\X^{AB}\,\,\equiv \Re
\Z^{AB}+i \Im \Z^{AB}, \ee \be \gY^A
 =Y^{A}+i\,\Y^{A}\equiv \Re \gY^{A}+i \Im \gY^{A}. \ee

 The real
part of $\Z^{\za\zb}$ is identified with   coordinates of the
 ten-dimensioned generalized space-time $X^{\za\zb}$ that contain in particular
Minkowski coordinates. The imaginary part
$\X^{\za\zb}=\Im\Z^{\za\zb}$ is required to be positive definite and
was treated in \cite{Mar} as a regulator
 that makes the Gaussian integrals well-defined
 (\ie physical quantities are obtained in the limit $\X^{\za\zb}\to
0$; note, that the complex coordinates $Z^{\za\zb}$ of \cite{Mar}
are related to $\Z^{\za\zb}$ via $ \Z^{AB} = i\bZ^{AB}$). The space
of coordinates $\Z^{\za\zb}$ forms the upper Siegel half-space
$\ZIGM$ \cite{Siegel,Igusa,Mumford}. Evidently,  $-\cZ^{AB}\in \ZIGM
$ provided that $\Z^{AB}\in \ZIGM $, and vice versa.

In \cite{gelcur} it was shown that complex conjugated holomorphic
and antiholomorphic in $\ZIGM$ solutions of the complexified
equations (\ref{dydy}) with $\mu=\pm i\hbar $
\be \label{dgydgyh+}
\left ( \f{\p}{\p \Z^{ A B}} + \,i\,\hbar \f{\p^2}{\p \gY^ A \p \gY^
B}\right ) C^+(\gY|\Z) =0\,, \ee \be \label{dgydgyh-} \left (
\f{\p}{\p \cZ^{ A B}} - \,i\,\hbar \f{\p^2}{\p \cY^ A \p \cY^
B}\right ) C^-(\cY|\cZ) =0\, \ee
\Big($ \overline{C^+(\gY|\Z)}=C^-
(\cY|\cZ) \Big)$ describe positive- and negative-frequency solutions of the massless
field equations. It should be noted that  $\ZIGM$ is  a $Sp(2M)$
invariant homogeneous space $Sp(2M|\mathbb{R})/ U(M)$ \cite{Igusa}.
A closely related intriguing and, most likely, important observation
 is that  solutions of the positive-frequency unfolded equations
(\ref{dgydgyh+})  with  $\dis{ \hbar
 = \f{1}{4 \pi  }
 }$ include Riemann theta-function \cite{gelcur}
\be
\label{genper}
C^+(\gY|\Z)= \sum_{n^A\in \mathbb{Z}^M} 
\exp{i(\pi\Z^{AB} n_A  n_B +2\pi n_C \gY^C)}\,,
\ee
which is the $D$-function
for the subclass of solutions periodic in the auxiliary variables
$\gY^A$. Indeed, Riemann theta-function is defined in the upper
Siegel half-space $\ZIGM$ and $Sp(2M|\mathbb{Z})$ plays the fundamental
role in its theory \cite{Igusa, Mumford}.
 This observation raises the question of the geometric interpretation
of the variables $\gY^A$ on the same footing as $\Z^{AB}$. This
question drives us immediately to the twistor theory
\cite{Bast_East,penr,ward} (also see \cite{witten}
and references therein for its applications in $N=4$ Yang-Mills
and string theory). One of the aims of this paper is to
clarify the $Sp(2M)$ invariant twistor construction
which in some important detail differs from the standard $SU(2,2)$
(\ie $4d$ conformal) setup \cite{penr,ward}, that corresponds to the case
of $M=4$. More generally, the original
motivation for this work was to elaborate a coordinate independent
global description of massless fields in Lagrangian Grassmannian.

In fact it is a folklore statement that there should be a deep
relation between unfolded dynamics and twistor theory, which, to
the best of our knowledge, has been never clearly spelled out so
far. In this paper we discuss this
correspondence in some more detail.
\renewcommand{\gn}{\nu}
\renewcommand{\gm}{\mu}
\renewcommand{\gc}{\chi}
\renewcommand{\gx}{\xi}
\renewcommand{\gf}{\zeta}
\renewcommand{\gr}{\rho}
\renewcommand{\ga}{\alpha}
\renewcommand{\gb}{\beta}

Indeed, in the problem of interest, it is self-suggestive that the
variables $Y$ \big(or $\gY$\big) are analogous to the twistor coordinates in
the usual twistor theory. $X^{AB}$ \big($\Z^{AB}$\big) are
coordinates of generalized (complexified) Minkowski space $
\M_M$ \big($\ZIGM$\big). Together,  $(X^{AB},Y^A)$ $\Big((\Z^{AB},\gY^A)\Big)$ are local
coordinates of the  correspondence space $\mathbf{C}$. The unfolded field
equation should encode the integral Penrose transform.
  Indeed,
generic solution of (\ref{dydy}) can locally be given in the form \cite{BHS}
\bee \label{genre}
 C(Y|X)
=\exp\left(-\mu X^{ A B}\f{\p^2}{\p Y^ A \p Y^ B }\right)C(Y|0)\,,
\eee
 where the ``initial data'' $C(Y|0)$ is an arbitrary function  of
$M$ variables $Y^ A$. Then, generic solution of massless field
equations (\ref{oscal}) and (\ref{ofer}) is determined by an
arbitrary (unrestricted) function $C(Y|0)$ on the twistor space $\mathbf{T}$
with local coordinates $Y$ according to (\ref{b0}) and (\ref{f0}).
These steps should provide a $Sp(2M)$ realization of a version of
Penrose transform,
expressed by the diagram\bee\label{diapen}
 \begin{picture}(200,80)( 0,26)
{\linethickness{.25mm}
\put(90,90){\vector( 1,-1){40}}%
\put(90,90){\vector( -1,-1){40}}%
\put(85,95)  {{    $\mathbf{C}$}}
\put(30,35)  {{   $\mathbf{M}$}}
\put(128,35)  {{   $\mathbf{T}\quad ,$}}
\put(55,70)  {{  \large $\eta$}}
\put(108,70)  {{  \large $\nu$}}
 }
\end{picture}
\eee
where $\mathbf{M} =Sp(2M )/P$  .

There is however important difference with the standard twistor
description of massless fields in usual (compactified) Minkowski
space. Indeed, the reduction of the equation (\ref{dydy}) to
Minkowski space is
\be \label{dydby}
\left ( \f{\p}{\p X^{\ga\da}}
+ \mu \f{\p^2}{\p y^\ga \p \bar{y}^\da}\right ) C(y|X) =0\,,
\ee
where $\ga,\gb\ldots 1,2$ and $\da,\db\ldots 1,2$ are two-component
indices with the convention that $A=(\ga,\da)$ and
$Y^A=(y^\ga,\bar{y}^\da)$. Apart from expressing auxiliary fields (\ie
 those depended both on $y^\ga$ and $\overline{y}^\da$) in
terms of space-time derivatives of the (anti)holomorphic fields,
the equation (\ref{dydby}) imposes the massless field equations on the latter
 \be \label{hol} \f{\p}{\p
y^{[\alpha}}\f{\p}{\p X^{\beta]\alpha^\prime}} C(y,0|X)=0 \,,\qquad
\f{\p}{\p \bar{y}^{[\alpha^\prime}}\f{\p}{\p X^{\alpha\beta^{\prime
]}}} C(0,\bar{y}|X)=0 \ee { and}
\be \label{kg4} \Box C(0,0|X)=0\,.
\ee

The equation (\ref{dydby}) is transformed to the first-order equation by a half-Fourier
transform with respect to $\bar{y}^\da$ such that
\be
\label{fourier}
\f{\p}{\p \bar{y}^\da}\longrightarrow i \pi_\da\,.
\ee In these terms the equation (\ref{dydby}) for the Fourier
transformed field $\widetilde{C}(y,\pi|X)$ takes the form
\be
\label{dydbyf} \left ( \f{\p}{\p X^{ \ga \da}} + \pi_\da \f{\p}{\p y^\ga
}\right ) \widetilde{C}(y,\pi|X) =0\,
\ee with the appropriate choice of
$\mu$.

Since
\bee\nn
 {C}(y,\bar{y}|X)
=\int   d^2 \pi\,\,\exp ( i\pi_\da \bar{y}^\da)
 \widetilde{{C}}(y,\pi|X)\,,
\eee
from (\ref{genre}) one has
\bee \label{genreP}
  {C}(y,\bar{y}|X)
&=&\int  d^2 \pi\,\,\exp (i\pi_\da \bar{y}^\da)
\exp\left(-i\mu X^{ \ga \db}\f{\p }{\p {y}^\ga   }\pi_\db\right)
  \widetilde{{C}}(y,\pi|0)\,
=\\ \nn
&=&\int   d^2 \pi\,\,\exp( i\pi_\da \bar{y}^\da)
  \widetilde{{C}}(y^\ga-i\mu X^{ \ga \db} \pi_\db,\pi|0)\,.
\eee
At $y=\bar y =0$ this formula reproduces the  nonprojective version of
the Penrose formula for massless fields \cite{penr}.

 Since  the Fourier transformed equation
(\ref{dydbyf}) is of first-order  it can be interpreted as the
condition
\be J C=0
\ee for some vector field $J$ on functions on
the correspondence space. This effectively reduces functions on the
correspondence space to those on the twistor space. The Fourier
transform (\ref{fourier}) distinguishes between   primed
and unprimed spinors in the twistor construction. Although somewhat
ugly, this is standard in twistor theory. In the case of
$\mathfrak{sp}(8)$ invariant equations  this trick  it is not possible.
Indeed, even after a half-Fourier transform at least some of the
unfolded equations in $\M_M$ contain second $Y$-derivatives,
$\dis{\f{\p^2}{\p y^\ga \p y^\gb}}$ or $\dis{\f{\p^2}{\p y^\da \p y^\db}}$.

Another important difference between the $Sp(8)$ invariant construction
of infinite  sets of higher-spin fields and the standard one used for
particular massless fields is that in the latter case  the helicity group
$U(1)\subset Sp(8)$  distinguishes between
different massless fields (\ie spins). The helicity generator is
\be
\Hh=y^\ga\f{\p}{\p y^\ga} - \by^\da\f{\p}{\p \by^\da}\,.
\ee
The centralizer of $\Hh$ in $\mathfrak{sp}(8)$ is
$\mathfrak{u}(2,2)=\mathfrak{su}(2,2)\oplus \mathfrak{u}(1)$, where
$\mathfrak{u}(1)$ is generated by $\Hh$.
Fields associated to a given spin have
definite homogeneity
\be \label{spin} \Hh C(Y) = \pm 2\,s \,C(Y)\,.
\ee
This means that the true coordinates appropriate for the
description of a massless field of definite helicity are of the
projective space invariant under the action of $\Hh$
\be
y^\ga\longrightarrow \exp(i\phi) y^\ga
\q
{\by}^\da\longrightarrow
\exp(-i\phi) \by^\da\,,
\ee equivalent to
\be
y^\ga\longrightarrow \exp(i\phi) y^\ga
\q
{\pi}_\da\longrightarrow
\exp(i\phi) {\pi}_\da\,.
\ee
In the case where all fields are involved, the condition (\ref{spin})
has to be relaxed. As a result, $(y^\ga, \pi^\da)$ become local (not
homogeneous) coordinates of the appropriate  twistor
space, as was realized by many authors
(see e.g. \cite{BLS1}). This has two different consequences.
The relevant symmetry group $G$ becomes not reductive and the twistor space
becomes noncompact. Simultaneously, the
system becomes $Sp(8)$ rather than $SU(2,2)$ symmetric.

In this paper we suggest a geometric realization of the
diagram (\ref{diapen}), that underlies the $Sp(8)$ invariant
systems, as well as   their $Sp(2M)$ generalizations.
The appropriate construction results from the coset space $SpH(2M)/P$.
Here $SpH$ is a semidirect product of $Sp(2M)$ with the Heisenberg
group $\HG$ with $2M$ noncentral elements (\ie $M$ pairs of
oscillators). As such it is not reductive.
 $P$ is some quasiparabolic subgroup of $SpH(2M)$.

\renewcommand{\gn}{F}
\renewcommand{\gm}{D}
\renewcommand{\gc}{{H}}
\renewcommand{\gx}{{Q}}
\renewcommand{\gf}{{F}}
\renewcommand{\gr}{{R}}
\renewcommand{\ga}{{A}}
\renewcommand{\gb}{{B}}

To achieve a coordinate independent description we find it
convenient to use the BRST language. Namely,   the quotient space
can be effectively described by
 the invariance condition
\be Q_P \,f =0\,,
\ee
where $Q_P$ is the canonical BRST operator of the
 group $P$. Indeed, for 0-forms $f$ this condition implies
that $f$ is constant on any orbit of $P$, \ie it is a function
on $SpH(2M)/P$. Since $Q_P$ is built from right vector fields of $P$ on
$SpH(2M)$, it is globally defined on $SpH(2M)$. The key observation
is that, in the local coordinates $X^{AB}$ and $Y^A$,
 the first term in Eq.(\ref{dydy}) can be interpreted as
the remaining right vector field $J_{AB}$ on $Sp(2M)$ while the
second terms amounts to the bilinear combination  $J_A J_B$ of the
remaining vector fields on the Heisenberg group. We therefore look for
a BRST operator $\QQ$ constructed in terms of $SpH(2M)$ right Lie
vector fields such that $\QQ^2=0$ and the condition $\QQ f=0$ gives
an extension of the equation (\ref{dydy}) to $SpH(2M)$. The construction of
this BRST operator is the main result of this paper, which provides
in particular the global description of the system.

The rest of the paper is organized as follows. We
start in Section \ref{Homogeneous manifolds} by recalling the construction of
Lagrangian Grassmannian  as a
quotient space. In Subsections \ref{HomRea}
and \ref{HomCom}, we consider, respectively,  the real
and complex cases. In Subsection \ref{HomCom}, the  realization of  the
 complex Siegel space as an $Sp(2M|\mathbb{R})$ orbit in
the complex Lagrangian Grassmannian is discussed.
In Subsection \ref{Heisen},
we recall the definition of real and complex Heisenberg group and introduce
higher rank Heisenberg extensions
of $Sp(2M|\mathbb{R} )$ and $Sp(2M|\mathbb{C} )$ as well as
higher rank Fock-Siegel spaces.
In Section \ref{CanBRST},
we recall some elementary facts on BRST operators.
In Section \ref{NonCanBRST}, higher rank
nonstandard $SpH(2M)$ invariant BRST operators $\QQ_r$
are introduced. In Section \ref{Vector fields},
 right invariant vector fields on $SpH_r(2M)$ are presented.
In Section \ref{UnfolBRST}, the higher--spin unfolded equations  are obtained
from $\QQ_1 f=0$.
Analogously, in Section \ref{cureq}, the higher--spin current equations are
obtained from  $\QQ_2 f=0$.

\section{Homogeneous manifolds}
\label{Homogeneous manifolds}
\subsection{Real Lagrangian Grassmannian}
\label{HomRea}
Let us recall the quotient space construction of Lagrangian Grassmannian.
 The group $Sp(2M|\mathbb{R})$ is
constituted by real matrices
\bee \label{sp2M}
\beee{c r } G= & \left( \beee{r r}
a^\ga{}_\gb&b^{\ga\gm}\\c_{C\gb}&d_C{}^\gm  \eeee \right)
 \eeee
{} \eee with $M\times M$ blocks
$a^\ga{}_\gb\,,\,b^{\ga\gb}\,,\,c_{\ga\gb}\,,\,d_\ga{}^\gb$ that
satisfy the relations \bee\label{sprel1}\!\!
a{}^{\ga}{}_{C }b{}^{{D}\,C}-
a{}^{D}{}_C b{}^\ga{}^C=0,\quad
 a{}^{\ga}{}_{C } d{}_B{}^{\,C}-b{}^\ga{}^C c{}_{ B \,C}=\gd^{\ga}{}_{ B}\,,
\quad   c{}_{B C} d{}_A{}^C- c{}_{A C} d{}_B{}^C   =0\,\, \eee
 equivalent to the invariance condition $AJA^t = J$
for the symplectic form
\be\label{skew} J= \left( \begin{array}{cc}
                   0    &   I^{\ga}{}_{\gb}   \\
                   -I_{C}{}^{D}   &   0
            \end{array}
    \right)\,,
\ee where $I$ is the  unit $M\times M$ matrix and $A^t$ is
a transposed matrix.
Note, that from (\ref{sprel1}) it follows that
\bee \label{sp2Mrev}
\beee{c r } G^{-1}= & \left( \beee{r r}
a^\prime{}^\ga{}_\gb&b^\prime{}^{\ga\gm}
\\c^\prime{}_{{C}\gb}&d^\prime{}_{C}{}^\gm  \eeee \right)
 \eeee
\beee{c r }  = & \left( \beee{r r}
d{}_\gb{}^\ga&-b^{\gm\ga}\\-c_{\gb{C}}&a{}^\gm{}_{C}  \eeee \right).
 \eeee
\eee

 $Sp(2M|\mathbb{R})$ contains the following important subgroups.
 The Abelian subgroup of translations  $\TT$ consists of  the elements
\bee t(X)= \left( \begin{array}{cc}
                   I    &   X   \\
                   0    &   I
            \end{array}
    \right)\,
\label{4} \eee with various  $X^{\ga\gb}=X^{\gb\ga}$. The product
law in $\TT$ is $ t(X)t(Y) = t(X+Y)$.

Analogously, the Abelian subgroup $\SSS$ of special conformal
transformations is constituted by the matrices (\ref{sp2M}) with $a=d=I$, $b=0$.
The subgroup $GL(M)$ of generalized Lorentz transformations $SL(M)$
and dilatations    consists of the matrices (\ref{sp2M}) with
$b=c=0$ and $a^{\gb}{}_{{C}} d_\ga{}^{C} =\delta_\ga^\gb\,.$

The parabolic subgroup $P(\mathbb{R}) $ of $Sp(2M|\mathbb{R})$ relevant to our
consideration is the closure of  $\SSS$ and $GL(M)$, \ie \be P \ni
p= \left( \begin{array}{cc}
                   a    &   0  \\
                   c   &   d
            \end{array}
    \right)\,.
\label{p} \ee It is a maximal parabolic subgroup. In notations of
\cite{Bast_East} it is depicted by the  diagram $\qquad
 \begin{picture}(120,4)(0,6)
{\linethickness{.25mm}
\put(30,9){\line(1,0){08}}%
\put(56,9){\line(1,0){05}}%
\put(41,8.5){\dots}%
 \put(0,9){\line(1,0){30}}%
\put(60,9){\line(1,0){30}}%
\put(90,8){\line(1,0){30}}%
\put(90,10){\line(1,0){30}}%
\put(111.5,5.5){{ \large $\mathbf \times$}} \put(100,5.5){{ \large
$\mathbf <$}}
 \put(-5,6.3)  {{  \bf $\bullet$}}
\put(25,6.3) {{ \bf $\bullet$}} \put(85,6.3) {{ \bf $\bullet$}}
\put(55,6.3) {{  \bf $\bullet$}} }
\end{picture}
\qquad$
 that results from crossing out the right node of the Dynkin
diagram  of $ {Sp}(2M|\mathbb{R})$.

Lagrangian Grassmannian $\C\M_M$ is the homogeneous  space
 \be \label{DC} \C\M_M = Sp(2M|\mathbb{R})/P(\mathbb{R}) \,,  \nn
\ee \ie it is constituted by the elements $h\in Sp(2M|\mathbb{R})$ identified
modulo the right action of $P{}(\mathbb{R})$ \be\nn h\sim h_1 = h p\,,\qquad
h\in Sp(2M|\mathbb{R})\,,\quad p\in P{(\mathbb{R})}\,. \ee

Any $G\in Sp(2M|\mathbb{R})$  with nondegenerate
 $d$ (\ref{sp2M}) can be represented in the form

\bee \label{spprod}
\beee{c c r r } \left( \beee{r r}
a^{A}{}_{B}&b^{{A}{C}}\\c_{{D}{B}}&d_{D}{}^{C}  \eeee \right)   =&
 \left( \beee{r r} \gd^{A}{}_{E}&  X{}^{A}{}^{F}
\\  0&\gd_{D}{}^{F}  \eeee \right)
& \left( \beee{r r} \A{}^{E}{}_{G}&\quad 0
\\\quad 0&\D_{F}{}^{H}  \eeee \right)
& \left( \beee{r r} \gd^{G}{}_{B}&0
\\ \C_{ H}{}_{B}&\gd_{H}{}^{C}  \eeee \right).
\eeee & \eee
This gives
\bee \nn \beee{  r r }
\left( \beee{r r}
a^{A}{}_{B}&b^{{A}{C}}\\c_{{D}{B}}&d_{D}{}^{C}  \eeee \right)   =&
 \left( \beee{r r} \A^A{}_{B}
 +X^{\ga {F}}\D_{F}{}^{G}\C_{{G}{B}}
   &\quad X^{\ga {F}}\D_{F}{}^{C}\\
    \D_{D}{}^{G}  \C_{{G}{B}}
   & \D_{D}{}^{C}   \eeee \right)\,,
\eeee \eee
where
\bee\label{coorsp}\A{}^{A}{}_\gb =(d^{-1}){}_\gb{}^{A}\,,\quad
X^{\ga\,\gm}
=b{}^\ga{}^{C} \A{}^\gm{}_{C}
, \quad \C_{{B}\,{A}}=c_{{C}{B}} \A{}^{C}{}_{A}
\eee can be chosen as local coordinates on $Sp(2M|\mathbb{R} )$,
where
$X^{\gb\,\ga}=X^{\ga\,\gb}$ and $\C_{\gb\,\ga}=\C_{\ga\,\gb}$ by
virtue of the identities
\bee\nn\quad
 - c{}_{\,{B}}{}_{{A}}d^{-1}{}_{{C} }{}^{B} +
c{}_{\,{B}}{}_{{C} }  d^{-1}{}_{A}{}^{B}=0\,,\quad  -b{}^\ga{}^{B}
d^{-1}{}_{B}{}^{C}+b{}^{{C}}{}^{\,{B}}d^{-1}{}_{{B}
}{}^{\ga}=0,
\eee
which follow from (\ref{sprel1}) and (\ref{sp2Mrev}).
Note that
$\D{}_{A}{}^{B}=d{}_{A}{}^{B}\,.
$

 The big cell $\M_M$ of $\C \M_M$ consists of the classes represented by
elements $t(X)$ of the group of translations $\TT$.
 $Sp(2M|\mathbb{R})$  acts in $\M_M$  by the
M$\ddot{\mathrm{o}}$bius transformation
 \bee \label{mob} \beee{  r r }
X^\prime = (AX+B )(CX+D)^{-1}\,\,  & ,  \quad   \quad
\left( \beee{r r} A&B\\
C&D  \eeee \right) \quad\in Sp(2M|\mathbb{{R}}).
 \eeee
\eee

By virtue of (\ref{spprod}) any element (\ref{sp2M}) of $Sp(2M|\mathbb{R})$ with $det
|d^\ga{}_\gb | \neq 0$ belongs to some equivalence class
associated to a point of  $\M_M$.
{}From (\ref{sprel1}), (\ref{p}) it
follows that $d$ is non-degenerate for any $p\in P(\mathbb{R}) $. As a result, the
 rank  of the  block $ d$ of a given element $g\in Sp(2M|\mathbb{R})$
(\ref{sp2M}) is the same for all  $g P{}$. Let $\rank(g)=
\rank\,|d|$  $\forall g\in \C \M_M$.
$\rank(g)$ characterizes different types of equivalence
classes, \ie different subsets of the compactified space-time
$\C\M_M$. Those with $\rank (g) < M$ correspond to generalized
conformal infinity.

\subsection{Complex Lagrangian Grassmannian and Siegel space}
\label{HomCom}
Complex Lagrangian Grassmannian\, $\C \M{}^{\mathbb{C}}_M=Sp(2M|\mathbb{C})/ P({\mathbb{C}})$\,
  results from
the analogous construction for $Sp(2M|\mathbb{C})$ and its complex
parabolic subgroup $ P({\mathbb{C}}) $. Local coordinates of
$\C \M{}^{\mathbb{C}}_M$ will be denoted by $Z.$
 The $Sp(2M|\mathbb{C})$
 M$\ddot{\mathrm{o}}$bius transformation has the same form (\ref{mob})
\bee \label{mobZ}
\beee{  r r } Z^\prime = (AZ+ B
)(CZ+D)^{-1}\,,\quad  &\qquad
\left( \beee{r r} A&B\\
C&D  \eeee \right) \quad\in Sp(2M| \mathbb{{C}}).
 \eeee
\eee
Being invariant under
$Sp(2M|\mathbb{R})\subset Sp(2M|\mathbb{C})$, $\C \M{}^{\mathbb{C}}_M$
is not $Sp(2M|\mathbb{R})$
homogeneous, containing different orbits.

 The following three orbits of $Sp(2M|\mathbb{R})$ are of most
 interest for us:

\bigskip

I.\,\,\,\, $\C \M_M$.

\bigskip

II.\,\,\, The
upper Siegel half-space $\ZIGM$ \cite{Siegel} of   matrices $\Z^{\ga\gb}$
with positive definite $ \Im \Z^{\ga\gb}$.

\bigskip

III. The lower Siegel half-space $\ZIGM{}^-$  of   matrices
$\Z^{\ga\gb}$ with negative definite $ \Im \Z^{\ga\gb}$.

\bigskip

Note, that both   $\ZIGM$ and  $\ZIGM{}^-$  can be realized as
$Sp(2M|\mathbb{R})/ U(M)$  \cite{Siegel,Igusa,Mumford}.

\subsection{Heisenberg extension}

\label{Heisen}

  The $(2M+1)-$dimensional Heisenberg group
$\HG =\mathbb{R}^M\times \mathbb{R}^M \times\mathbb{R}^1$
constituted by \bee\label{Heis} \FFf=\{\ff \,,\,u\}\qquad
\ff=y^\ga\,,\,w_\gb\, \qquad \ga,\,\,\gb=1,...,M \eee
has the product law
$$\FFf_1\circ\,\FFf_2=\{\ff_1{}+\ff_2{} \,,\, \,u_1+u_2-
( \ff_1\,, \, \ff_2)\}\,,$$ where  $(\,,\,)$ is the symplectic form
\bee\label{simfor} ( \ff_1\,, \,\ff_2)
=y_1{}^\ga\,\,w_2{}{}_\ga-y_2{}^\ga\,\,w_1{}_\ga\,=-( \ff_2\,,
\,\ff_1)\,, \qquad \ga = 1,\ldots, M .
\eee
The Heisenberg group $\HG(\mathbb{R})$
contains two quasiparabolic subgroups $\HG^\pm(\mathbb{R})$
\bee\label{qusiparab} H{}^-
=\{\,0\,,\,w_\ga\, \,,\,u \},\quad \quad H{}^+ =\{ y^\ga\,,\,0\,\,,\,u \}.
 \eee
 The respective quotient spaces $\HG \,/H^\pm$ are $R{}^\pm\sim \mathbb{R}^M$.
Note that the appropriate spaces of functions on $R{}^\pm  $ form Fock spaces.
Complexification of this construction is straightforward,
$\HGC =\mathbb{C}^M\times \mathbb{C}^M \times\mathbb{C}^1$, \etc.

  $Sp(2M|\mathbb{R})$ \Big($Sp(2M|\mathbb{C})$\Big) acts canonically on  $\HG$
\Big($\HGC$\Big) which is the manifestation of the standard fact that
$Sp(2M)$ possesses the oscillator realization (see e.g. \cite{BG}).
This makes it possible to introduce a semi-direct product $SpH(2M ) =Sp(2M ) \htimes \HG$
   \bee\label{Sph}
SpH(2M )\,:\,\G=\{G \,,\,\FFf  \}\,,\qquad 
G\in Sp(2M),\quad \FFf=\{\ff \,,\,u\}\in  \HG \eee
with the product law
\be\nn \G_1\circ \G_2=\{ G_1 G_2 \,,\,\,\,
\ff_1\,+\,G_1  \ff_2\, \,,\,\,\,u_1\,+\,u_2\,-\,(\ff_1\,,G_1 \ff_2)\}\,, \ee
where $(\,,\,)$ is the symplectic form (\ref{simfor}) and
\bee\nn \beee{c r l r c } G   \ff= &\ls \left(\! \beee{r r}
                         a^\ga{}_\gb&b^{\ga\gm}\\c_{{C}\gb}&d_{C}{}^\gm
                          \eeee \!\right)
&
 \ls\left(\!\! \beee{r  }
y{}^\gb \\w_{\gm}   \eeee\!\! \right)=
&\ls\, \left( \!\beee{r }
              a^\ga{}_\gb y{}^\gb+ b^{\ga\gm}w_{\gm}
              \\
               c_{{C}\gb}y{}^\gb+d_{C}{}^\gm w_{\gm}
          \eeee \!\!\right),
&\!\!
  G \in Sp(2M),\,\,\,  \ff =(y^\ga,w_\gb)\in\HG.
\eeee \eee
Analogously, over any field $\mathbb{A}$ and for any natural $r$ we
 introduce a rank $ r$
Heisenberg extensions   $ SpH_r(2M |\mathbb{A})$ as
   \bee\label{semi}
SpH_r(2M|\mathbb{A} )=Sp(2M|\mathbb{A}) \htimes
\underbrace{
\HG(\mathbb{A})\times\dots\times
\HG(\mathbb{A})}_r
 .\eee
  Note, that $SpH_1(2M )\equiv SpH(2M )$.
When it does not lead to misunderstandings, we will use  shorthand
 notation like  $SpH$  instead of $SpH(2M|\mathbb{R})$ or $SpH (2M|\mathbb{C})$
 and $PH$  instead of $PH(2M|\mathbb{R})$ or $PH(2M|\mathbb{C})$ \etc.

 Consider the  lower quasiparabolic  subgroup $\PH(2M|\mathbb{R})=P{}\htimes H^-
 \subset SpH(2M|\mathbb{R})$
 \bee \nn\label{parab}
\beee{c r c } \PH (2M|\mathbb{R})  =\left\{\rule{0pt}{18pt}\right.& \left( \beee{c c}
p^\ga{}_\gb&0\,\\  p_{C}{}_\gb&p_{C}{}^D  \eeee \right),
&0\,,\,p_\ga\,,\,p\left.\rule{0pt}{18pt}\right\}\,.
\eeee
\eee
Local coordinates on   $SpH/ \PH $ can be chosen as \,\,
\be  X^{\ga\gb}\,\,
,\quad Y^\ga=y^\ga- w_{\gb}X^{\ga\gb} \label{XY}.\ee
Analogously we define $SpH_r/ \PH_r$ with local coordinates $
X^{\ga\gb}\,$ and\,\, $Y_1{}^\ga,\dots,Y_r{}^\ga$\,, and their
complexifications   with
local coordinates $  Z^{\ga\gb}\,$ and\,\,
$\gY_1{}^\ga,\dots,\gY_r{}^\ga$\,.

The orbit of $Sp(2M|\mathbb{R})$ in
$ SpH(2M|\mathbb{C})/  \PH(2M|\mathbb{C})$ with   positive
definite $ \Im Z^{\ga\gb}$  is the upper
Fock-Siegel space $\GZM$ introduced
in \cite{gelcur}.  The orbit with negative definite $ \Im Z^{\ga\gb}$
is the lower Fock-Siegel space $\GZM^-$.
Consider a space  $U$ of functions\footnote{Using this abused
terminology to simplify the presentation, we assume that functions should be
substituted by appropriate sections of fiber bundles whenever necessary.}
 $C(\G)$ on $SpH/\PH$ valued in a one dimensional $PH(2M|\mathbb{R})$-module
 $V$. Following standard induced module construction, require
\be \label{fact} 
C(\G\circ \P )= \det{}^\gga\big( p^A{}_B\big)
\exp (-i\half \hbar  p)\, C(\G)
,\qquad \G\in SpH/\PH,\quad \P\in  \PH,\ee
where the generalized conformal weight $\gga$ and Plank constant $\hbar$ 
 are arbitrary parameters that characterize $V$. The space $U$ forms a left
 $SpH(2M|\mathbb{R})$-module with  the transformation law
\be \nn 
C(\G){\longrightarrow}C_\B (\G)= C(\B\circ\G)\,,\qquad \B\in SpH(2M|\mathbb{R}).
\ee
In local coordinates $X$, $Y$
(\ref{XY})
 for $\B$ of the form (\ref{sp2M}) one obtains,
 \be
 \label{tr}
 C_\B (Y|X) =  \det{}^{-\gga} |cX+d|
 \exp( -\half i\hbar  Y^{\prime\,\,\ga} Y^\gb c_{\ga\gb} )\,C(Y^\prime|X^\prime )\,,
 \ee
 where
 \be\nn
 X^\prime = (aX+b )(cX+d)^{-1}\,,\qquad Y^\prime = (cX+d)^{-1} Y\,.
 \ee
For $\dis{\gga=\half}$\,, (\ref{tr}) is  the $Sp(2M)$ transformation law for solutions
 of the equations (\ref{dydy}) with
   $\mu=\f{i}{2\hbar}$, infinitesimal form of which was obtained in  \cite{Mar}.
 This suggests that it should be possible to impose
  the equation (\ref{dydy}) in a manifestly $Sp(2M)-$invariant and coordinate
  independent way.
  This  will be achieved in Section \ref{NonCanBRST} using BRST technics.

Analogously, the space  of complex functions $\mathbf{f}(\G)$
on the upper Fock-Siegel space $\GZM
\subset
 SpH(2M|\mathbb{C})/ \PH(2M|\mathbb{C})$  valued in a one dimensional
$PH(2M|\mathbb{C})$ module subjected to (\ref{fact}) forms a $SpH(2M)$-module
under the group action (\ref{tr}).
For $\B\in Sp(2M|\mathbb{R})$ of the form (\ref{sp2M})  one obtains
in local coordinates $\Z$, $\gY$
 \be
 \label{trC}
 C_\B (\gY{}|\Z{}) = C(\gY{}^\prime|\Z{}^\prime ) \det{}^{-\gga} |c\Z{}+d|
 \exp(  -i\half\hbar\gY{}^{\prime\,\,\ga} \gY{}^\gb c_{\ga\gb} )\,,
 \ee
 where
 \be\label{sptrZY}
 \Z{}^\prime = (a\Z{}+b )(c\Z{}+d)^{-1}\,,\qquad \gY{}^\prime = (c\Z{}+d)^{-1} \gY{}\,.
 \ee
 This formula with $\gga=\half$ is in agrement with the
 fact that Riemann theta functions solve the complexified unfolded
 higher--spin field equations (\ref{dydy}) with $\dis{\mu = \f{i}{4 \pi  } }$.
\section{Canonical BRST operator}
\label{CanBRST}
\newcommand{\FF}{F}
Any Lie group $G$ possesses two mutually commuting sets of  left
 and right Lie
vector fields $J^l_{B}$ and $J^r_{A}$ (${B}\,,\,\,{A}=1,2,\ldots,
\dim G$), each forming Lie algebra $\mathfrak{g}$ of $G$
\be\nn
[J_{A}^{r}\,,J_{B}^r ]=f_{{A}{B}}{}^{E} J_{E}^r\q
[J_{A}^{l}\,,J_{B}^l ]=f_{{A}{B}}{}^{E} J_{E}^l\q
[J_{A}^{r}\,,J_{B}^l ]=0\,. \ee Let
$I^r_\ga$ be a subset of
$J_{A}^{r}$ that corresponds to some subgroup $B\subset G$. The
space  of functions on $G/B$ identifies with the space
of functions on $G$ that satisfy
\be \label{rinv} I^r_\ga \FF(G) =0\,.
\ee
The algebra Lie $\mathfrak{g}$ acts on solutions of (\ref{rinv}) by the
left Lie vector fields $J_\gm^{l}$.

An important extension of (\ref{rinv})  to the
induced modules construction  is
\be \label{ind} (I^r_\ga+T_\ga) \FF(G) =0\,,
\ee where $\FF(G)$ is valued in some $B$-module $V$
and $T_\ga$ provide a representation of the Lie algebra
$\mathfrak{b}$ of $B$ on $V$,
\be [T_\ga\,,T_\gb ]=f_{\ga\gb}{}^{C}
T_{C}\q [J_B^{r}\,,T_\ga ]=0\q [J_B^{l}\,,T_\ga ]=0\,.
\ee
In
what follows we will be interested in the particular case of this
construction with a  one dimensional $B$--module $V$. In
this case, $\FF(G)$ is still valued in $\mathbb{R}$ or $\mathbb{C}$
and $T_\ga$ is given by some constants associated to central
elements of the grade zero  part of   $\mathfrak{b}$.

A useful way to impose the condition (\ref{ind}) is by introducing a
canonical BRST (Chevalley-Eilenberg) operator
\be\label{CheEil}
  Q = c^\ga
(I^r_\ga + T_\ga)- \half c^\ga c^\gb b_{C} f_{\ga\gb}{}^{C}\q
 Q^2  =0\,,
\ee where the ghosts $c^\ga$ and $b_\ga$
obey the Clifford anticommutation relations
\be
\{c^\ga\,,b_\gb\}=\delta^\ga_\gb\q \{c^\ga\,,c^\gb\}=0\q
\{b_\ga\,,b_\gb\}=0\,.
\ee
The equation (\ref{ind}) results from the restriction
to the sector of $c$--independent $\FF(G)$ of the condition
\be \label{QQ} Q  \FF(G,c) =0\,,
\ee
where $b_\ga$ is realized as
\be b_\ga =\f{\p}{\p c^\ga}\,.
\ee
 The BRST construction provides the
extension of the equation (\ref{ind}) to the case of
$\FF(G,c)$ that belongs
to the Clifford-Fock module generated from the vacuum that satisfies
$b_\ga|0\rangle=0$.
This extension is expected to have useful applications in
the context of the so-called $\sigma_-$ cohomology in
the unfolded dynamics approach \cite{SV} (see also \cite{BHS,tens2}
and \cite{solv} for a review) as well as for the analysis of conserved
charges by BRST methods.

The standard twistor theory diagram relates three objects: the
correspondence space $\mathbf{C}$, space-time $\mathbf{M}$ and twistor space $\mathbf{T}$. In
the standard twistor theory all of them are homogeneous spaces.
In the model of interest this is not
quite the case. Although the spaces $\mathbf{C}$ and $\mathbf{M}$ indeed allow the
quotient space realization with
\bee\label{quotSM}
\mathbf{C}=SpH(2M|\mathbb{C})/\PH(\mathbb{C})\,,\quad \mathbf{M}= \C\M_M^{\mathbb{C}}=
SpH(2M|\mathbb{C})/\big(P( \mathbb{C}){}\htimes \HG(\mathbb{C})\big) \,, \eee
 this is not so for the twistor space $\mathbf{T}$ that should have
$Y^A$ as local coordinates. Indeed, it is not hard to see that no
appropriate $Sp(2M)$ homogeneous space with local coordinates $Y^A$
exists. The point is that the realization of $Sp(2M)$ on functions
of the coordinates $Y^A$, that can be read of the equation
(\ref{dydy}) is given by a
second-order differential operator that cannot
result from a first-order vector field. It turns out however that
appropriate spaces of functions on all three spaces $\mathbf{C},\,\,\,\mathbf{M}$ and $\mathbf{T}$
can be described by the BRST conditions $Q f=0$. In the cases of
$\mathbf{C}$ and $\mathbf{M}$ the $Q_\PH$ and $Q_{P\hltimes H}$
are canonical for the groups $\PH$ and $P{}
\htimes \HG$, respectively.

Namely, according  to (\ref{CheEil}),
\bee\ls \label{QP} Q_\PH=
   c^\ga{}_\gb (J^\gb{}_\ga+ \gga \gd^\gb{}_\ga)  +
c_{\ga\gb} J^{\ga\gb} +c (J-\hhhh )+ c_\ga J^\ga \\ \nn
+c^\ga{}_\gb c^\gb{}_{C} b^{C}_\ga -2c^\ga{}_\gb
c_{\ga{C}}b^{\gb{C}} -c_\ga{}^\gb c_\gb b^\ga\,,
\eee
 where
\be J^\ga{}_\gb\q J^{\ga\gb}\q J_{\ga\gb}\q J^\ga\q J_\ga\q J\,  \ee
are the right vector fields  of  $SpH$,
while    constants   $ \hhhh $
 and  $\gga $
characterize the induced   $ {\PH}$-module in question.
 The operator $Q_{P \hltimes H}$ is
 \bee\label{QPH} Q_{P\hltimes H}=
 c^\ga{}_\gb (J^\gb{}_\ga+ \gga \gd^\gb{}_\ga)  +
c_{\ga\gb} J^{\ga\gb} +c J + c_\ga J^\ga
+ c^\ga J_\ga
\\ \nn
+c^\ga{}_\gb c^\gb{}_{C} b^{C}_\ga -2c^\ga{}_\gb
c_{\ga{C}}b^{\gb{C}} -c_\ga{}^\gb c_\gb b^\ga 
+c^\ga{}  c {}_\ga b+c_{\ga{B}}c^{\ga}b^{{B}}
+c_\ga{}^\gb c^\ga b_\gb   .\quad \eee
The both are nilpotent,
 \bee\nn Q^2_{P\hltimes H} =
Q^2_{\PH}=0\,, \eee
as a consequence of the (anti)commutation relations
\bee\nn [J^\ga{}_\gb \,,J^{C}{}_{E}] &=& \delta_\gb^{C}
J^\ga{}_{E} -\delta_{E}^\ga J^{C}{}_\gb\,,
\\\nn
[J^\ga{}_\gb \,,J^{{C}{E}}] &=& \delta_\gb^{C} J^{\ga{E}} +
\delta_\gb^{E} J^{\ga{C}}\,,
\\   \label{comrelsp}
[J^\ga{}_\gb \,,J_{{C}{E}}] &=& -\delta^\ga_{C} J_{\gb{E}} -
\delta^\ga_{E} J_{\gb{C}}\,,
\\  \nn
[J_{\ga\gb}\,,J^{{C}{E}} ] &=& \delta_\ga^{C} J^{E}{}_\gb+
\delta_\gb^{C} J^{E}{}_\ga +\delta_\ga^{E} J^{C}{}_\gb+
\delta_\gb^{E} J^{C}{}_\ga \,,
\eee
\bee \nn [J^\ga{}_\gb
\,,J^{{C}}] &=& \delta_\gb^{C} J^{\ga}\q\qquad\,\,\quad
[J^\ga{}_\gb \,,J_{{C}}] = -\delta^\ga_{C} J_{\gb}\,,
\\    \label{comrelsph}
[J_{\ga\gb}\,,J^{{C}} ] &=&
 \delta_\gb^{C} J_\ga{}+\delta_\ga^{C} J_\gb{}\q
[J^{\ga\gb}\,,J_{{C}} ] =
 -\delta^\gb_{C} J^\ga{}-\delta^\gb_{C} J^\gb{}\,,
\\  \nn
[J_\ga\,,J^\gb ] &=& \delta_\ga^\gb J \,
\eee
and
\bee \nn \{c^{\ga\gb} \,,b_{{C}{E}}\} =\half( \delta^\ga_{E}
\delta^\gb_{C}+ \delta^\gb_{E} \delta^\ga_{C} )\,, \quad
\{c_{\ga\gb} \,,b^{{C}{E}}\} =\half(
\delta_\ga^{E} \delta_\gb^{C}+ \delta_\gb^{E} \delta_\ga^{C}
)\,,
\\ \nn\{c^\ga{}_\gb
\,,b^{C}{}_{E}\} = \delta^\ga_{E} \delta_\gb^{C}\,,\quad
\{c^{\ga} \,,b_{{C}}\} = \delta^\ga_\gb ,\quad \{c_{\ga}
\,,b^{{C}}\} = \delta_\ga^\gb ,\quad \nn \{c \,,b\} = 1. \eee
(Other (anti)commutation relations are zero.)

\section{Nonstandard BRST operator}
\label{NonCanBRST}

The main observation of this paper is that there exists a
nonstandard BRST operator $\QQ$ that is built from the right
Lie vector fields and supplements the quotient conditions (\ref{QQ})
with the equations (\ref{dydy}). Since left Lie vector fields commute
to the right ones,
the resulting equations  are manifestly invariant under the
left action of $\mathfrak{g}$ at any point of $G$. Hence
this construction is $G$ invariant and coordinate independent.

 $\QQ$ has the form
\be\label{fullbrstlow} \QQ=Q_\PH+\triangle \QQ \,.
\ee
Here $Q_\PH$ is the
standard BRST operator (\ref{QP}) of the parabolic subalgebra which
reduces $SpH$ to the correspondence space. The additional part
\bee\label{triangQ} \ls\triangle \QQ=
c^{\ga\gb}(J_{\ga\gb} \!- \!\hhh J_\ga J_\gb)
\!+\!2 c^\ga{}_\gb c^{\gb{C}}b_{\ga{C}} 
\!-\!4 c^{\ga\gb} c_{\gb{C}} b^{C}{}_\ga
\!-\!2\hhh c^{\ga\gb} c_{\ga\gb}b\,
\!+\quad
\\ \nn
+2\hhh c^{\ga\gb}c_\gb b J_\ga\!-4\hhh c^{\ga{C}} c_{{C}\gb} c_\ga b
b^{\gb} + 4\hhh  c^{\gb {C}} c_{\gb\ga} b^\ga J_{C} - 4\hhh c^{\gb
{C}} c_{\gb\ga} c_{{C}{E}} b^{E} b^\ga\, .\quad \eee
 takes care of the coordinates $X^{AB}$, ensuring that
the space $\mathbf{T}$ has local coordinates $Y^A$.

The main result of this paper is that $ \QQ^2=0 $
 provided that the generalized conformal weight is
\be
\label{gamma}
\dis{ \gga=\half\,.}
\ee

To explain our construction, let us first consider the compatibility of
the conditions $P_\A f=0$ for the set of operators
\bee \label{genB}\!
P_\A:\quad \!\!P^{A}{}_{B}\!=\! J^{A}{}_{B}\!+\! \gamma \gd^{A}{}_{B},\quad
\!P^{\ga\gb}\!=\!
J^{\ga\gb} \!,\quad \!{P^\ga\!=\! J^\ga}\!,\quad
\!P\!=\! J-\hhhh
 ,\quad
  \! P_{\ga\gb}\!= \! J_{\ga\gb} +\kappa J_\ga J_\gb ,\,
 \eee
where $\gamma$, $\nu$ and  $\kappa$ are free parameters to be determined.
Namely, we demand that the operators $P_\A$ (\ref{genB})  form
a ``closed algebra"
\bee\label{brack}
[P_{\A}, P_\Bb]=\phi_{\A\Bb}^\C P_\C
\eee
with   some
``structure functions" $\phi_{\A\mathcal{B}}^\C$ that may depend on the
vector fields $J$.

Using (\ref{comrelsp}) and (\ref{comrelsph}) it is easy to see that
the only commutation relations
that do not obey (\ref{brack}) for generic values of the parameters
 $\gamma$, $\nu$ and  $\kappa$  are
\bee\nn
[P_{\ga\gb},\,P^{C }\,]&=&\kappa (\gd_\ga^{C} J_\gb+\gd_\gb^{C} J_\ga)
(\kappa^{-1}+ \hhhh+ P )
\,, \\ \nn
 [P_{\ga\gb},P^{\gm\gn}]&=&\gd_\ga^\gm P_\gb{}^\gn+\gd_\gb^\gm
P_\ga{}^\gn +\gd_\ga^\gn  P_\gb{}^\gm+\gd_\gb^\gn P_\ga{}^\gm
- \kappa
(\gd_\ga^\gm \gd_\gb{}^\gn+\gd_\gb^\gm \gd_\ga{}^\gn )P
\\ \nn
&+&\kappa(\gd_\ga^\gm J_\gb P{}^\gn+\gd_\gb^\gm J_\ga P{}^\gn
+\gd_\ga^\gn  J_\gb P{}^\gm+\gd_\gb^\gn J_\ga P{}^\gm )\\ \nn
&-&(2\gamma+\kappa\nu) (\gd_\ga^\gm \gd_\gb{}^\gn+\gd_\gb^\gm \gd_\ga{}^\gn )\,
  . \eee
Hence, $P_\A$ obey (\ref{brack}) at the condition
(\ref{gamma}) along with
\be \label{munu}
  \kappa =-\nu^{-1} \,.
\ee
 In this case, the nonzero commutation relations
between  $P_\A$ take the form
 \bee\nn
[P_{\ga\gb},P^{\gm\gn}]&=&\,\,\gd_\ga^\gm P_\gb{}^\gn+\gd_\gb^\gm
P_\ga{}^\gn +\gd_\ga^\gn  P_\gb{}^\gm+\gd_\gb^\gn P_\ga{}^\gm
+\hhh
(\gd_\ga^\gm \gd_\gb{}^\gn+\gd_\gb^\gm \gd_\ga{}^\gn )\,P\\ \nn
& &-\hhh(\gd_\ga^\gm J_\gb +\gd_\gb^\gm J_\ga )P{}^\gn
-\hhh(\gd_\ga^\gn  J_\gb  +\gd_\gb^\gn J_\ga) P{}^\gm \,,
\\\nn 
[P^\ga{}_\gb \,,P^{C}{}_{E}] &=& \delta_\gb^{C} P^\ga{}_{E}
-\delta_{E}^\ga P^{C}{}_\gb\,,
\\ \nn 
[P^\ga{}_\gb \,,P^{{C}{E}}]&=&
\delta_\gb^{C} P^{\ga{E}} + \delta_\gb^{E} P^{\ga{C}}\,,
\\ \label{comrelB}
[P_{\ga\gb} ,P^D{}_{C}]&=& \gd_\ga^D P_{\gb{C}}+\gd_\gb^D
P_{\ga{C}}
\,,
\\\nn 
[P_{\ga\gb},\,P^{C }\,]&=&-\nu^{-1}(\gd_\ga^{C} J_\gb+\gd_\gb^{C}
J_\ga)P,
\\ \nn
[P^{\ga}{}_{\gb} ,P^D ]&=&  \gd_\gb^D P^{\ga}\,. \eee

Since some of the right hand sides of the relations (\ref{comrelB}) contain products of $J$
and $P$, that do not commute according to (\ref{genB})\,, (\ref{comrelsp})
and (\ref{comrelsph}), the relations (\ref{comrelB}) form a nonlinear algebra.
Naively, one might think that $\nu^{-1}$ is a deformation parameter,
which describes the nonlinear algebra (\ref{comrelB}) as a deformation
of a Lie algebra at $\nu^{-1}=0$. This is not true,
however, because of the relation
\be\nn
[J_A,\, P^B ]=\gd_A^B(P+\nu)\,,
\ee
that has to be used in the consistency check. The terms, that differ
the relations (\ref{comrelB}) from a Lie algebra, include all $\nu$-dependent terms
along with the $\nu$-independent right-hand side of the last of the relations
(\ref{comrelB}). On the other hand, at $\nu^{-1}=0$, the operators
$ P^{A}{}_{B}$, $ P^{\ga\gb}$, $P_{\ga\gb}$ and $P$
 form a  Lie algebra $sp(2M)\oplus u(1)$. This implies that
 the  BRST operator
\bee\label{qsp}
Q_{sp}&=& c^\ga{}_\gb P^\gb{}_\ga +c^{\ga\gb}P_{\ga\gb} +
c_{\ga\gb} P^{\ga\gb} +c P+\\ \nn
&+&c^\ga{}_\gb c^\gb{}_{C} b^{C}_\ga -2c^\ga{}_\gb
c_{\ga{C}}b^{\gb{C}}
+ 2 c^\ga{}_\gb c^{\gb{C}}b_{\ga{C}}
-4 c^{\ga\gb} c_{\gb{C}} b^{C}{}_\ga
 \eee
squares to zero at $\nu^{-1}=0$, which observation simplifies the
 computation of $\QQ^2$ sketched below.

That $P_\A$ satisfy (\ref{brack}) allows us to look for a nilpotent operator of the form
\be
\label{genq}
\QQ=c^\A P_\A
- \half \sum_{n>0}  \phi_{\A_1 \dots \A_n\A_{n+1}}^{\Bb_1  \dots \Bb_{n}}(J)\,
 c^{\A_1} \dots c^{\A_n} c^{\A_{n+1}}\,b_{\Bb_1} \dots b_{\Bb_{n}}\,,
\ee
where $\phi_{\A\gb}^{\C}$ are the ``structure functions" of (\ref{brack}).
The higher structure functions that enter the terms of higher orders in
ghost fields can appear in the case of field-dependent
$\phi_{\A\gb}^{\C}$ as was first found in the Hamiltonian analysis of supergravity
\cite{fv}. A general analysis for  classical Hamiltonian systems
 was given in \cite{ff}. The extension to the quantum case of associative algebras,
 which is of interest for us here, was considered in \cite{bf}.

The Jacobi identity expresses the identity $[Q_0\,,Q_0^2]=0$ for $Q_0=c^\A P_\A$.
In the case of a field-dependent structure function $\phi_{\A\mathcal{B}}^\C$
it has the form
\be
\label{jac}\Big (
c^\A c^\Bb c^\C [P_\C \,, \phi_{\A\Bb}^\D ]
+c^\A c^\Bb c^\C    \phi^\E_{\A\Bb}\,\phi^\D_{\C \E} \Big )
P_\D=0\,.
\ee
Differently from the case of constant $\phi_{\A \mathcal{B}}^\C$, the condition
(\ref{jac}) does not
imply that the expression in brackets is zero. Instead, it imposes a weaker condition
\be
\label{jac1}c^\A c^\Bb c^\C \Big(
[P_\C \,, \phi_{\A\Bb}^\D ]  +  \phi^\E_{\A\Bb}\,\phi^\D_{\C \E}
+   2\phi_{\A\Bb\C}^{\D\E}P_\E + \phi_{\A\Bb\C}^{\F\E} \phi_{\F\E}^\D
\Big )=0\,,
\ee
where $\phi_{\A\Bb\C}^{\D\E}=-\phi_{\A\Bb\C}^{\E\D}$ are some new structure coefficients
that, in turn, should be determined from (\ref{jac1}).
Indeed, one can see
that (\ref{jac}) follows from (\ref{jac1}).  Provided that (\ref{jac}) is true, the operator
\be\nn
\QQ=Q_0+Q_1+Q_2\q
\ee
where
\be\nn
Q_0= c^\A P_\A \q Q_1= -\half
c^\A c^\Bb \phi_{\A\Bb}^\C b_\C\q
Q_2=-\half c^\A c^\Bb c^\C \phi_{\A\Bb\C}^{\D\F}b_\D b_\F\,,
\ee
is nilpotent up to the terms
$
\QQ^2 =O(c^4 b^2)\,.
$
Generally, these terms are compensated by the terms with higher structure coefficients
in (\ref{genq}). Fortunately, in the case of interest, the terms of higher orders in $c$
and $b$ in $\QQ$ (\ref{genq}) are not needed because the $c^4 b^2$-type terms cancel out.

To check  nilpotency of $\QQ$, one starts with $Q_0+Q_1$
that accounts for all terms linear and quadratic in $c^\A$
\bee\nn
Q_0+Q_1=
Q_{sp}+
 c_\ga P^\ga 
  -c_\ga{}^\gb c_\gb b^\ga
-2\hhh c^{\ga\gb} c_{\ga\gb}b\,
+ 4\hhh  c^{\ga C}c_{\ga\gb} b^\gb J_C
+2\hhh c^{\ga\gb}c_\gb b J_\ga\,,
\eee
where $Q_{sp}$ is given in (\ref{qsp}).
Taking into account that $Q_{sp}^2$ only contains terms that are zero at $\nu^{-1}=0$,
it is not hard to obtain
\be\nn
(Q_0+Q_1)^2=4\hhh c_{\ga\gb}c^{\ga C}(
-c_C b P^\gb-2c_{C D} b^D P^\gb+ c_C  b^\gb P)\,.
\ee
These terms are cancelled by
\bee\nn
Q_2 =-4\hhh(  c^{\ga\gb} c_{\gb{C}} c_\ga b   b^{{C}}
+ c^{{A} {C}} c_{{A}{E}} c_{{C}{B}}b^{E} b^{B})\,.
 \eee
Namely,
 \be\nn
 \{ Q_0+Q_1,\,Q_2\}=-(Q_0+Q_1)^2\,,
 \ee
 and
\bee\nn
(Q_2)^2=
   (4\hhhh)^{-2}c_{{D}{E}} c^{{D} {F}}
  c_{{F}{\ga}} c^{\ga\gb} c_{\gb{C}}    b^{E}  b^{{C}} b
 =  0
\eee
because the tensor  \,$F_{EC}=c_{{D}{E}} c^{{D} {F}}
  c_{{F}{\ga}} c^{\ga\gb} c_{\gb{C}} $\, is symmetric in the indices $E$ and $C$.

Let us comment on the relation of our construction with
the BRST operators that follow from the constraint algebra of
twistor particle models considered for example in
\cite{shir,BC,BL,BLS1,BHS}. Roughly speaking, the latter correspond
to the first term of the operator (\ref{triangQ}) \ie to those terms
that only contain the ghost $c^{AB}$. The extension
to its Heisenberg algebra counterpart was discussed in \cite{FE1}.
Reductions of $\QQ$ of this type  trivially square to zero.
Usually such operators are given in  particular coordinates like
$X^{AB}$ and $Y^A$. The operator
$\QQ$ constructed in this paper extends this construction in two important
respects. Firstly, it extends  the construction to full $Sp(2M)$.
This in particular determines  the generalized $Sp(2M)$ conformal
weight $\gamma$  of the massless fields in $\M_M$.
Secondly, it is formulated in a coordinate-independent way and is globally
defined thus giving rise to the coordinate independent version of $Sp(2M)$
invariant unfolded equations (\ref{dydy}). Indeed, to work in any coordinate
system it suffices to substitute  the corresponding expressions for the right
Lie vector fields into $\QQ$.

Otherwise the proposed BRST operator differs
from most versions of the BRST constructions used in
 higher--spin theory (see e.g. \cite{AB,ST,FT,Barnich:2004cr,BBPT,BG1,BLS}
 and references therein). In our approach,
we do not introduce {\it ad hoc} any auxiliary space and constraints to guess a
BRST operator that gives rise to appropriate dynamics, working
directly in terms of the symmetry group and its Lie vector fields,
that makes the setting manifestly symmetric, coordinate independent
and globally defined.
Let us stress that it is impossible to introduce a deformation  parameter into
$\QQ$ to treat it as a deformation of a BRST operator associated to some Lie algebra.
(Note that there is some similarity with the
BRST description of higher spins in $AdS$ background \cite{ST},
 where however, the BRST operator is a deformation of the standard one
 in Minkowski space.)

  The generalization of\, $\QQ$ \,(\ref{fullbrstlow})\, to the case
of the rank \,$r$ \,    quasiparabolic\, subgroup \\
$P \htimes \underbrace{\HG{}^-\times\dots\times\HG{}^-}_r $\,
  is straightforward
  \bee\label{fullbrstr}\nn
 \ls\QQ_{r}&=&
   c^\ga{}_\gb P^\gb{}_\ga   +
c_{\ga\gm} P^{\ga\gm}
+ c^{\ga\gb}P_{\ga\gb}+\sum_{j=1}^r
\big( c_j
P{}_{j} + c{}_{j}{}_\ga P{}_{j}{}^\ga\big)\,\,\,\,\\ \nn &&
+c^\ga{}_\gb c^\gb{}_{C} b^{C}_\ga
-2c^\ga{}_\gb c_{\ga{C}}b^{\gb{C}}
+2 c^\ga{}_\gb c^{\gb{C}}b_{\ga{C}}
-4 c^{\ga\gb} c_{\gb{C}} b^{C}{}_\ga\,\,\,\,\\ \nn &&
 - \sum_{j=1}^r
 c_\ga{}^\gb c{}_{j}{}_\gb b{}_{j}{}^\ga
 + \sum_{j=1}^r\! \hhh_j\Big(\!2 c^{\ga\gb} c_{\ga\gb}b_j \!
+\!2 c^{\ga\gb}c_j{}_\gb b_j J_j{}_\ga
+ 4   c^{\ga
\gb} c_{\ga C} b_j{}^C J_j{}_\gb \,\,\,\,\\ \nn&&
- 4  c^{\ga\gb} c_{\gb C} c_j{}_\ga b_j   b_j{}^{C} - 4  c^{\ga B} c_{\ga C}
c_{B{E}}  b_j{}^C b_j{}^{E}\Big)\,,\,\,\,
\eee
where
\bee\nn  \label{genBr}
P_\A:\quad P^{A}{}_{B}&=&
J^{A}{}_{B}+  \gamma_r \gd^{A}{}_{B} ,\quad
P^{\gm{B}}= J^{\gm{B}} , \quad  P_i^{A} =
J_i^{A} ,\quad \nn P_i{}=
J_i{}-\hhhh_i{} \\ \quad P_{\ga\gb}&=&   J_{\ga\gb} + \sum_{i=1}^r \kappa_i
J_i{}{}_\ga J_i{}{}_\gb
\eee
and
\be\label{munugar} \kappa_i =-\nu_i{}^{-1} \q
 \gamma_r=\frac{r}{2}\,. \ee
Nonzero
anticommutation relations of the Clifford ghosts are
\be
\label{anticomrelr} \{c_j{}^{\ga} \,,b_k{}_{{\gb}}\} =
\delta^\ga_\gb\gd_j{}_k \,,\quad \nn \{c_j{}_{\ga} \,,b_k{}^{{\gb}}\}
= \delta_\ga^\gb\gd_j{}_k \,,\quad \nn \{c_j{} \,,b_k{}\} =
\gd_j{}_k\,.
\ee
One can make sure that
$
\big(\QQ_{r}\big)^2=0. $
Indeed, using that
$[J_i{}{}_\gb ,J_j{}{}{}^\gm]=0$ at $i\neq j\,,$
it is easy to see that operators (\ref{genBr})
 form a ``closed algebra" with nonzero commutation relations
\bee \nn 
[P^\ga{}_\gb \,,P^{C}{}_{E}] &=& \delta_\gb^{C}
P^\ga{}_{E} -\delta_{E}^\ga P^{C}{}_\gb
\,,\\ \nn [P^\ga{}_\gb
\,,P^{{C}{E}}]&=& \delta_\gb^{C} P^{\ga{E}} + \delta_\gb^{E}
P^{\ga{C}}\,,
\\ \nn
[P_{\ga\gb},P^{\gm\gn}]&=&\gd_\ga^\gm P_\gb{}^\gn+\gd_\gb^\gm
P_\ga{}^\gn +\gd_\ga^\gn  P_\gb{}^\gm+\gd_\gb^\gn P_\ga{}^\gm
+
\sum_j\hhh_j P_j{}                                     \nn
(\gd_\ga^\gm \gd_\gb{}^\gn+\gd_\gb^\gm \gd_\ga{}^\gn )\,\\ \nn
&-&\sum_j\hhh_j(\gd_\ga^\gm J_j{}{}_\gb J_j{}{}{}^\gn+\gd_\gb^\gm J_j{}{}_\ga
J_j{}{}{}^\gn +\gd_\ga^\gn  J_j{}{}_\gb J_j{}{}{}^\gm+\gd_\gb^\gn
J_j{}{}_\ga J_j{}{}{}^\gm )
\,,\\\nn
[P_{\ga\gb} ,P^D{}_{C}]&=& \gd_\ga^D P_{\gb{C}}+\gd_\ga^D
P_{\ga{C}}\,\\ \nn 
[P_{\ga\gb},P_j^{C }]&=&(\gd_\ga^{C} J_j{}_\gb+\gd_\gb^{C} J_j{}_\ga)P_j
\,,
\\\nn 
[P^{\ga}_{\gb} ,P_j^D ]&=&  \gd_\gb^D
P_j^{\ga}. \eee
provided that (\ref{munugar}) is true.
The rest of the analysis
is identical to  the case of rank one.

\section{Vector fields on $SpH(2M)$}
\label{Vector fields}
\subsection{Rank $1$   }

To make the equations explicit in one or another coordinate system,
a concrete realization of the right Lie vector fields is needed.
Straightforward computation  in the  local coordinates
   \be\label{coorsph}\A_\gb{}^B\,,\quad X^{\ga\,\gb}\,,
 \quad \C_{\ga\,\gb}\,,\quad y^\ga\,,\quad w_\ga\,,\quad u\,
 \ee
of  $SpH(2M|\mathbb{R}) $ gives
 \bee \label{vect_fields sp}
J_{\ga\,\gb}&=&
 -2 \A{}^{ {E} }{}_{\ga}\A{}^{{D} }{}_{\gb}\f{\p}{\p X^{{D}{E}}}
+   2\A{}^{{E}}{}_{(\gb}\C_{ \ga){D} }\f{\p}{\p \A{}^{{E}}{}_{{D}}}
+2\C_{\ga{D}}\C_{\gb{E}}\f{\p}{\p \C_{{D} {E}}} ,\rule{18pt}{0pt}\\ \nn
J^{A\,B}&=& 2 \f{\p}{\p \C_{A\,B}},\\ \nn
J_A{}^{B}&=&-2\C_{A{C}}\f{\p}{\p \C_{B{C}}}
               -\A{}^{{C}}{}_{A}\f{\p}{\p \A{}^{{C}}{}_{B}},
\eee
 \bee\label{vect_fields gz}
J^{{C}}&=&\D{}_{B}{}^{C}\Big( X{}^{B}{}^{A}
 \f{\p}{\p y^{{A}}} +\f{\p}{\p w_{{B}}} + \Big(-
y^{{B}}+
 w_{{A}}  X{}^{B}{}^{A}
\Big)\f{\p}{\p u}\Big),
\\ \nn
 J_{\ga}&=&
-\A{}^{B}{}_\ga \Big(
\f{\p}{\p y^{{B}}}+w_{{B}}\f{\p}{\p u}\Big) -\C{}_\ga{}_B  J^B,
\\ \nn
J&=&2\f{\p}{\p u}\,.
\eee
\subsection{Rank $r$  }
  $SpH_r(2M)$   right
vector fields consist of the $Sp(2M )$ right  vector
fields (\ref{vect_fields sp}) and $r$ mutually commutative copies
 of the  vector fields (\ref{vect_fields gz})
\bee\label{vect_fields gzr}
J_{j}{}^{{C}}&=&\D{}_\gm{}^{C}\Big(
X{}^\gm{}^{A}   \f{\p}{\p y_{j}{}^{{A}}} +\f{\p}{\p w_{j}{}_{\gm}}+\Big(-
y_{j}{}^{\gm}+
 w_{j}{}_{{A}}  X{}^\gm{}^{A}
\Big)\f{\p}{\p u_{j}{}}\Big),
\\ \nn
 J_{j}{}_{\ga}&=&
- \A{}^\gm{}_\ga\Big(
\f{\p}{\p y_{j}{}^{\gm}}   +
 w_{j}{}_{\gm}  \f{\p}{\p u_{j}{}}\Big)
-\C{}_\ga{}_\gn J_{j}{}^{\gn},
\\ \nn
J_{j}{}&=&2\f{\p}{\p u_{j}{}}, \eee where $j=1,...,r.$

  Let us consider more closely the case of rank two, that will be used
in Section \ref{cureq} to construct $Sp(2M)$ invariant {\it current
equations}.
Introducing
 \bee \nn J_\pm=\half(J_1\pm J_2),\quad y_\pm=y_1\pm y_2,\quad w_\pm=(w_1\pm w_2),
 \quad u_\pm=(u_1\pm u_2),
\eee
 we obtain from (\ref{vect_fields gzr})
 \bee\label{vect_fields gz-F}
J_{-}{}^{{C}}&=&\D{}_\gm{}^{C}\Big(\f{\p}{\p w_{-}{}_{\gm}} +  X{}^\gm{}^{A}
 \f{\p}{\p y_-{}^A}
-\half Y_{+}{}^{\gm} \f{\p}{\p u_{-}{}}
 -\half Y_-{}^{\gm} \f{\p}{\p u_{+}{}}
\Big),
\\ \nn
 J_{-}{}_{\ga}&=&
- \A{}^\gm{}_\ga \Big( \f{\p}{\p y_-{}^{\gm}}
 +\half w_{-}{}_{\gm} \f{\p}{\p u_{+}{}}+
\half w_{+}{}_{\gm} \f{\p}{\p u_{-}{}}\Big)
  -\C{}_\ga{}_\gn J_{-}{}^{\gn},
\\ \nn
J_{-}{}&=&2\f{\p}{\p u_{-}{}}, \eee

\bee\label{vect_fields gz+F}
J_{+}{}^{{C}}&=&\D{}_\gm{}^{C}\Big(\f{\p}{\p w_{+}{}_{\gm}} +  X{}^\gm{}^{A}
  \f{\p}{\p y_{+}{}^{{A}}}
-\half Y_{+}{}^{\gm} \f{\p}{\p u_{+}{}}
-\half Y_{-}{}^{\gm} \f{\p}{\p u_{-}{}}\Big)
,
\\ \nn
 J_{+}{}_{\ga}&=&
- \A{}^\gm{}_\ga\Big(
\f{\p}{\p y_{+}{}^{\gm}}
+\half  w_{+}{}_{\gm}  \f{\p}{\p u_{+}{}}
+\half   w_{-}{}_{\gm}  \f{\p}{\p u_{-}{}}\Big)
-\C{}_\ga{}_D J_{+}{}^D,
\\ \nn
J_{+}{}&=&2\f{\p}{\p u_{+}{}}, \eee
where
$Y_{\pm}{}^{\gm}=y_{\pm}{}^{\gm}-
 w_{\pm}{}_{{A}}  X{}^\gm{}^{A}$
.

\section{BRST operator and unfolded equations }
\label{UnfolBRST}
Let $f$ be a function on
 $SpH/\PH$, independent
 of the Clifford elements $c$ and $b$. Let $\QQ$ be the BRST operator
(\ref{fullbrstlow}). Then the condition
\be\label{0brstunfol}
\QQ f=0
\ee
implies \bee\label{brsteq1}
     (J^\gb{}_\ga+ \half \gd^\gb{}_\ga)f=0,
    \\ \label{brsteq2}   J^{\gn\gm}f=0,
    \\ \label{brsteq3}      (J-\hhhh )f=0,
    \\ \label{brsteq4}        J^\gr f=0,
    \\ \label{brsteq5}       (J_{\ga\gb} - \hhh J_\ga J_\gb)f=0.
\eee

Substituting (\ref{vect_fields sp}) and (\ref{vect_fields gz}) into
 (\ref{brsteq1})-(\ref{brsteq4})
we obtain
 \bee\label{rank1brsteq1}
\Big(
               -\A{}^{{C}}{}_{B}\f{\p}{\p \A{}^{{C}}{}_{D}}
+\half\gd_B^{D} \Big)f=0\,,
    \\ \label{rank1brsteq2}
   \f{\p}{\p \C_{A\,B}}f=0\,,
    \\ \label{rank1brsteq3}
     \Big(-2\f{\p}{\p u}+\hhhh \Big)f=0\,,
     \\\label{rank1brsteq4}
       \Big(  \f{\p}{\p w_{{A}}}
+  X{}^{B}{}^{A}  \f{\p}{\p y^{{B}}} +\Big(-      y^{{A}}+
 w_{{B}}  X{}^{B}{}^{A}
\Big)\f{\p}{\p u} \Big) f=0\,.
   \eee
Note, that Eqs. (\ref{rank1brsteq1})-(\ref{rank1brsteq4}), that
correspond to the subgroup $\PH\subset SpH$, are first order differential equations
with respect to $\A$,
$\C$, $u$ and $w$, respectively,  thus, reconstructing
 the dependence on the coordinates of $\PH$. Hence we can set $w=0$ , $ \C=0  $ in (\ref{brsteq5}).

Substituting (\ref{vect_fields sp}),(\ref{vect_fields gz}),(\ref{rank1brsteq2}) and (\ref{rank1brsteq3})
into (\ref{brsteq5}) and taking into account
 (\ref{brsteq4})  ,  we obtain
 at $w=0$ , $ \C=0  $ in   local coordinates $X$, $Y$ (\ref{XY})
  \bee\label{1brstunfol}
 \Big( \f{\p}{\p X^{A\,B}}+\half
  \hhh
\f{\p}{\p Y^{A}} \f{\p}{\p Y^{B}} \Big)f  =0\,,
\eee
which is  Eq.(\ref{dydy})   at $\mu=\half\hhh $.

 Since by virtue of (\ref{comrelsph}) and (\ref{brsteq3})
$[J_\ga\,,J^\gb]=\hhhh\gd_\ga^\gb $, we set
\be\label{hbar}\hhhh=-i\hbar .
\ee
\big(Note that the normalization of (\ref{dgydgyh+})  results
 from (\ref{1brstunfol}) via the rescaling
  $Y^{A}\rightarrow \frac{1}{\sqrt{2}\, \hbar} Y^{A}$.\big)

Since Eq.(\ref{1brstunfol}) is of
first order in  $X^{\ga\gb}$,
it  reconstructs the dependence on  $X$, \ie  solutions of
(\ref{0brstunfol})   are parametrized by functions   $f(Y)$
 on  the twistor space.

General solution of (\ref{0brstunfol}) is
 \bee\label{ungensol}f= \det(\A)^{\half}\exp
 \Big(\half \hhhh \big(u +w_{B} Y^{B}\big)\Big)   f_0  (Y|X  ), \eee where $
f_0{}(Y|X )$ is any solution of
(\ref{1brstunfol})\,.

Proceeding analogously with complex vector fields on $SpH(2M|\mathbb{C})$,
 we obtain complex    unfolded equations
\bee\label{2brstunfol}
 \Big( \f{\p}{\p Z^{A\,B}}+ \half  \hhh
\f{\p}{\p \gY^{A}} \f{\p}{\p \gY^{B}} \Big)f =0\,.
\eee

As  mentioned in Section \ref{Homogeneous manifolds}, a natural complexification of generalized
space-time $\M_M$ is the upper Fock-Siegel space $\GZM\subset
 SpH(2M|\mathbb{C})/ \PH(2M|\mathbb{C})$.
For $Z=\Z\in \ZIGM $
 Eqs.~(\ref{2brstunfol}) coincide  with the  field
 equations for   massless
fields in the
  Fock-Siegel space obtained in \cite{gelcur}
  up to a coefficient in front of the second term.

\section{Rank 2  BRST operator and current equations}
\label{cureq} $\mathfrak{sp}(2M|\mathbb{R})$ invariant current equations introduced in
\cite{gelcur} to formulate HS charge conservation in $\M_M$ have
the form
\be \label{unfol2_Fur} \left( \, \f{\p}{\p \Z{}^{AB}}  +
 \,\hbar\, \W_{(A}
 \f{\p}{\p   \gY{}{\,}{}^{B)}}
 \right) F =0\,.
\ee
 Let us explain how they result from  the
nonstandard BRST operator construction. Using the rank $r$ BRST
operator (\ref{fullbrstr}) for $r=2$ one can write the current
equations in the coordinate independent $SpH(2M)$ invariant form.

Setting %
$\hhhh_1{} =\hhhh{} $ and $\hhhh_2{}=-\hhhh{}$
  and using notations
$c_\pm=c_1\pm c_2$ and $J_\pm=\half(J_1\pm J_2)$ we obtain \bee
\label{brst2} {\QQ}_2\big|_{b=0}=
   c^\ga{}_\gb (J^\gb{}_\ga+   \gd^\gb{}_\ga) +
   c^{\ga\gb}(J_{\ga\gb} - 4\hhh J_+{}_\ga J_-{}_\gb) +
c_{\ga\gb} J^{\ga\gb} +\\ \nn
 +c_+ J_+ +c_-( J_--\hhhh )
+c_+{}_A J_+{}^A +c_-{}_A J_-{}^A
.\eee
So, for a rank $2$ field $F(\A,\C,X,y_\pm,w_\pm,u_\pm)$ independent of the
Clifford elements $c_\pm$ and $b_\pm$, the condition
$$
\QQ_{2} F=0
$$
implies
 \bee
  \label{brst20}   J_+ F=0\,,
\\\label{brst21}   ( J_--\hhhh )F=0\,,
\\ \label{brst22}    (J^\gb{}_\ga+   \gd^\gb{}_\ga)F=0\,,
\\\label{brst23}     J^{\ga\gb}F=0\,,
\\\label{brst24}  J_+{}^\ga F=0\,,
\\ \label{brst25}  J_-{}^\ga F=0\,,
\\\label{brst26}    (J_{\ga\gb} -
4\hhh J_{-}{}_{(\gb}J_{+}{}_{\ga)}
)F=0\,.
 \eee Substituting the
expressions (\ref{vect_fields sp}), (\ref{vect_fields gz-F}) and
(\ref{vect_fields gz+F}) to  (\ref{brst20})-(\ref{brst23}) we obtain
 \bee
 \label{Jpm}
 \f{\p}{\p u_{+}}F&=&0,\\
 \label{uhalfh}
\Big(\f{\p}{\p u_{-}}- \half \hhhh \Big)F&=&0,\\
\label{C1}
 \Big(\A{}^{{C}}{}_{A}\f{\p}{\p \A{}^{{C}}{}_{B}}-\gd{}^{B}_{A}\Big)F&=&0,
 \\
 \label{C0}
 2 \f{\p}{\p \C_{\ga\gb}}F&=&0
 .
\eee
By virtue of (\ref{C0})  we can set $\C=0$. So, using  (\ref{Jpm})  we obtain
 from (\ref{brst24}) and (\ref{brst25})
\bee
 \label{Jpm2}
\Big(\f{\p}{\p w_{+}{}_{\gm}} +  X{}^\gm{}^{A}
  \f{\p}{\p y_{+}{}^{{A}}}
 +\f{1}{4}\hhhh\, Y_{-}{}^{\gm}   \Big)F=0
  ,
\\
\label{Jpm1}
\Big(  \f{\p}{\p w_{-}{}_{{A}}} +  X{}^{A}{}^{B} \f{\p}{\p y_{-}{}^{B}}
 +\f{1}{4}\hhhh\, Y_{+}{}^{A}
\Big) F=0.
\eee

Since the equations (\ref{Jpm1}), (\ref{Jpm2}) reconstruct the
dependence on $w_\pm$, we   can  set $w_\pm=0$ in (\ref{brst26}),
 whence by
virtue of (\ref{C0}),   (\ref{Jpm}), (\ref{vect_fields gz-F}) and
(\ref{vect_fields gz+F})  we have
 \bee \label{CurYY}
 \Big(\f{\p}{\p X^{AB}}
+2 \hhh\,\, \f{\p}{\p Y_{-}{}^{(B}}
\,\f{\p}{\p Y_{+}{}^{A)}}
 \Big) F=0\,.
\eee
Setting
$$ \f{\p}{\p W_+{}}=Y_-\qquad
W_+=-\f{\p}{\p Y_-{}} \,,$$
which substitution is analogous
to the Fourier transform (\ref{fourier}),
we obtain from (\ref{CurYY})
\bee \label{Cur}
 \Big(\f{\p}{\p X^{AB}}
-2 \hhh\,\, W_{+}{}_{(B}\,\f{\p}{\p Y_{+}{}^{A)}}
 \Big) F=0\,.
\eee

Analogously for the  complex version of the vector fields
(\ref{vect_fields sp}), (\ref{vect_fields gz-F}) and
(\ref{vect_fields gz+F}) we  obtain \bee \label{CurZ}
  \Big(\f{\p}{\p \Z^{AB}}
- 2\hhh\,\, {\mathcal{W} }_{+}{}_{(B}\,\f{\p}{\p \gY_{+}{}^{A)}}
 \Big) F=0\,.
\eee
Up to notations and  coefficients this gives the current equations   (\ref{unfol2_Fur}).

\section{Unfolded dynamics and twistors}
\label{UnfoTwist}

In this section we touch very briefly some general aspects of the
analogy between the unfolded dynamics approach and twistors.

Let $M^d$ be a $d$ dimensional manifold with coordinates $x^\un$
($\un = 0,1,\ldots d-1$). By unfolded formulation of a linear or
nonlinear system of differential equations in $M^d$ we mean its
equivalent reformulation in the first-order form
 \be \label{unf} dW^\Phi (x)= G^\Phi
(W(x))\,,
\ee where $\dis{ d=dx^\un \frac{\p}{\p x^\un}\, }$ is the
exterior differential on $M^d$, $W^\Phi(x)$ is a set of degree
$p_\Phi$-differential forms and $G^\Phi (W)$ is some degree $p_\Phi
+1$ function of the differential forms $W^\Phi$ \be\nn G^\Phi (W) =
\sum_{n=1}^\infty f^\Phi{}_{\Omega_1\ldots \Omega_n}
W^{\Omega_1}\wedge \ldots \wedge W^{\Omega_n}\,, \ee where the
coefficients $f^\Phi{}_{\Omega_1\ldots \Omega_n}$ satisfy the
(anti)symmetry condition
$
f^\Phi{}_{\Omega_1\ldots\Omega_k\Omega_{k+1} \ldots \Omega_n} =
(-1)^{p_{\Omega_{k+1}}p_{\Omega_k}}
f^\Phi{}_{\Omega_1\ldots\Omega_{k+1}\Omega_k \ldots  \Omega_n} \,
$
(extension to the supersymmetric case with an additional
boson-fermion grading is straightforward) and $G^\Phi$ satisfies the
condition \be \label{BI} G^\Omega (W)\wedge \f{\p G^\Phi (W)} {\p
W^\Omega}  =0\, \ee (the derivative $\frac{\p}{\p W^\Omega}$ is
left) equivalent to the
 generalized Jacobi identity on the structure coefficients
\be \label{jid} \sum_{n=0}^{m} (n+1) f^\Lambda{}_{[\Phi_1 \ldots
\Phi_{m-n+1}}  f^\Phi{}_{\Lambda\Phi_{m-n} \ldots \Phi_m\}} =0\,,
\ee where the brackets $[\,\}$ denote appropriate
(anti)symmetrization of indices $\Phi_i$. Given solution of
(\ref{jid}) it defines a
 free differential algebra  \cite{FDA1,FDA2}.

This method of describing dynamical systems was originally proposed
in  \cite{4dun,Ann} where it was applied to the analysis of free and
interacting massless  gauge fields
 in four dimensional anti-de Sitter space. The name
 {\it unfolded formulation} was given somewhat later  \cite{un}.
The method turned out to be very efficient and was further developed
in a number of papers (see e.g. \cite{act,solv,33,BIS} for recent
discussions).

Let us note that, to some extent, unfolded formulation is analogous to
the Cartan prolongation approach with the important difference
however that it is extended to dynamical fields that are
differential forms. This is crucial in several
respects. In particular, in this approach geometry is described by
differential forms via vielbein one form (ladder form) and Lorentz
connection. The same time, differential forms are gauge fields
analogous to vector potential in spin one Maxwell-Yang-Mills theory
or vielbein in spin two gravity theory. The presence of gauge fields
 is crucial for interacting (\ie nonlinear) theories. In addition, the
exterior algebra formalism makes the unfolded equations coordinate
independent and manifestly gauge invariant (in the latter case
provided that the system is universal \cite{act,solv}. Note that in
this case the unfolded system amounts \cite{act,BG} to some
$L_\infty$ algebra \cite{Linf}).

An important property of the unfolded dynamics is that, in the
topologically trivial situation, degrees of freedom are concentrated
in zero-forms $\go^i_0(x_0)$ at any $x=x_0$. This is a consequence
of the Poincare' lemma: the unfolded equations express all exterior
derivatives in terms of the values of fields themselves modulo exact
forms that can be gauged away by the gauge transformations.
Locally, what is left is the ``constant part" of the zero-forms.

This simple observation has a consequence that, to describe a system
with an infinite number of degrees of freedom like a massless field,
it is necessary to work with an infinite set of zero-forms that form an
infinite dimensional module of the space-time symmetry $\mathfrak{g}$
($\mathfrak{g}=\mathfrak{sp}(8)$, $\mathfrak{su}(2,2)$, $\mathfrak{iso}(1,3)$ {\etc}). In fact,
the module carried by zero-forms turns out to be dual (complex
equivalent)  to the space of single-particle states in the
respective quantum field theory \cite{BHS}
 which is the theory of massless fields of all
spins in the case of most interest in this paper.

Usually, infinite sets of zero-forms are realized as a space of
functions on some auxiliary space $\mathbf{T}$ $  \grave{a} \,\,\,{ la}$ (\ref{gener}).
Particular examples are provided by the ``generalized Weyl tensors"
$C(Y|X)$ or $C(y,\bar{y}|x)$ discussed in Section
\ref{Generalities}. In the gauge theory of higher-spin fields they describe
gauge invariant field strengths built  from gauge connection one-forms
(for more detail see \cite{33} and references therein).

$\mathbf{T}$ is an analogue of the twistor space in twistor theory. The most
important difference with twistor theory is that, in the unfolded
dynamics approach, the symmetry $G$  is not necessarily geometrically
realized in $\mathbf{T}$ (\ie by vector fields at the infinitesimal level).
Typically, $\mathbf{T}$ is realized as a Fock module where $G$ acts via
embedding into the Weyl algebra of oscillators. In particular,
$G=Sp(8)$ acts just this way in $\mathbf{T}$. This usually leads to the
appearance of operators nonlinear in the annihilation operators (\ie
  $\f{\p}{\p Y^A}$ in our case) that act by higher-order
differential operators hence driving us away from the usual twistor
setup.

In general, two types of models may appear upon unfolding. The
\emph{effective} type is that where the forms $W^\Lambda$ in the
unfolded dynamics turn out to be unrestricted in the twistor space
$\mathbf{T}$, or restricted by simple homogeneity conditions, eventually
implying that $\mathbf{T}$ is a projective space as is the case if one
considers a field of definite spin like in the standard twistor
theory. The \emph{ineffective} type is that where the $0$-forms
responsible for local degrees of freedom are not arbitrary in an
appropriate space $\mathbf{T}$ being themselves described as solutions of
differential conditions in $\mathbf{T}$ that may be as complicated as the
original field equations.

In the effective case the unfolded field equations map arbitrary
functions on $\mathbf{T}$ to solutions of the field equations on space-time
$\mathbf{M}$, thus describing a Penrose transform. In the ineffective case,
the unfolded field equations are still useful in many respects (in
particular for introducing nonlinear field interactions in a
coordinate-independent way \cite{non}), but may be not particularly
helpful for finding their explicit solutions.

There are two classes of models in higher-spin theory. The vector-like, that
work in any dimension \cite{non} are ineffective in the sense that
the structure of zero-forms is fairly complicated requiring
reductions and factorizations of ideals in certain noncommutative
spaces. For example, in \cite{act} it was shown that, in the
 vector-like approach, Einstein and Yang-Mills equations
are unfolded in such a way that the corresponding zero-forms should
themselves satisfy the Einstein and Yang-Mills equations in $\mathbf{T}$.

The spinor-like (or twistor-like) models are effective, operating
with zero-forms valued in unrestricted spaces of spinor variables
like $Y^A$. So far, these models have been formulated at the
nonlinear level only in three and four space-time dimensions (see
\cite{Gol} and references therein). However, as has been argued in
\cite{BHS,Mar,BPST} models of this class  are likely to allow
extension to the $Sp(2M)$ case considered in this paper and, as a
result, to higher dimensions including $d=6,10,11$.

\section{Conclusions}

We hope that this paper sheds light on the relation between
unfolded dynamics and twistor theory. The unfolded
formulation is somewhat less restrictive, which means of course that
some of the methods of twistor theory may not be directly extended
to unfolded dynamics. For example, we have shown that the
twistor-like description of $Sp(8)$ invariant field equations for
massless fields requires a nontrivial extension of the standard
twistor approach. Interestingly enough, the current equation
proposed in \cite{gelcur}, which ensures the current conservation in
$Sp(8)$ invariant field theories, belongs to the normal twistor case
of first order equations whose geometric meaning consists of the
factorization of the correspondence space to the twistor space.

The parallelism between unfolded dynamics and twistor theory goes
far enough. In particular, the unfolded dynamics approach
effectively reformulates dynamical field equations in terms of
$\mathfrak{g}$-modules where $\mathfrak{g}$ is a Lie algebra where
one-forms take their
values (for more detail see \cite{act,solv,33} and references
therein). This definitely has a lot of similarity with the general
approach of \cite{Bast_East}. In particular,  dynamical content of
unfolded equations (independent fields, invariant differential
operators, etc) is computed in terms of the so-called $\sigma_-$
cohomology \cite{SV}. (For more detail see \cite{BHS,solv,tens2}.)
For example, the $Sp(8)$ invariant equations (\ref{oscal}) and
(\ref{ofer}) were derived by this method in \cite{BHS}. We expect
that $\sigma_-$ cohomology should be related to the sheaf
cohomology in twistor theory. We hope to analyze this and  other
questions on the interplay between unfolded dynamics and twistor
theory in the future.

Another interesting problem for the future is to establish a
formal correspondence of the proposed BRST construction
with the unfolded equations which should in some sense
be dual to each other. In particular, the unfolded formulation
is also globally defined once left invariant forms on the corresponding
group manifold $SpH(2M)$ are given. The latter however can be directly
 read of the Lie vector fields on $SpH(2M)$. This suggests that
 there should be a direct way to relate the two approaches.
 Hopefully, the BRST approach proposed in
 \cite{Barnich:2004cr,BG1} may be useful in this respect.

 \section*{Acknowledgments}
This research was supported in part by INTAS Grant No 05-7928,
 RFBR Grant No 08-02-00963, LSS No 1615.2008.2
M.V. acknowledges a partial support from the Alexander von Humboldt
Foundation Grant PHYS0167.

\addtocounter{section}{1} \addcontentsline{toc}{section}{
\,\,\,\,\,\,References}
\medskip

\section*{$\rule{0pt}{1pt}$}

\end{document}